\documentclass{article}
\usepackage{spconf,amsmath,graphicx}
\usepackage{marvosym}
\usepackage{spconf,graphicx,tabularray,multirow}
\usepackage{hyperref}
\usepackage{enumitem}
\usepackage{verbatim}

\setlist{nosep, leftmargin=14pt}
\usepackage{multirow}
\usepackage{mwe} 
\usepackage{makecell}

\title{BS-Diff: Effective Bone Suppression\\
Using Conditional Diffusion Models from Chest X-Ray Images}

\name{Zhanghao Chen$^1$, Yifei Sun$^1$, Ruiquan Ge$^{1,8,*}$,  Wenjian Qin$^2$,Cheng Pan$^3$, Wenming Deng$^4$, 
\\ Zhou Liu$^4$, Wenwen Min$^5$, Ahmed Elazab$^6$, Xiang Wan$^{7}$, Changmiao Wang$^{7,*}$
\thanks{*Correspondings: gespring@hdu.edu.cn, cmwangalbert@gmail.com.}}

\address{$^1$Hangzhou Dianzi University, Hangzhou, China \\
$^2$Shenzhen Institute of Advanced Technology, Chinese Academy of Sciences, Shenzhen, China \\
$^3$Sanda University, Shanghai, China \\
$^4$National Cancer Center, Chinese Academy of Medical Sciences and Peking Union Medical College, \\
National Clinical Research Center for Cancer, Cancer Hospital \& Shenzhen Hospital, Shenzhen, China  \\
$^5$Yunnan University, Kunming, China   $\,$
$^6$Shenzhen University, Shenzhen, China  \\
$^7$Shenzhen Research Institute of Big Data, China $\,$
$^8$Hangzhou Institute of Advanced Technology, China \\
}

\begin{document}
%
\maketitle
\begin{abstract}
Chest X-rays (CXRs) are commonly utilized as a low-dose modality for lung screening. Nonetheless, the efficacy of CXRs is somewhat impeded, given that approximately 75\% of the lung area overlaps with bone, which in turn hampers the detection and diagnosis of diseases. As a remedial measure, bone suppression techniques have been introduced. The current dual-energy subtraction imaging technique in the clinic requires costly equipment and subjects being exposed to high radiation. To circumvent these issues, deep learning-based image generation algorithms have been proposed. However, existing methods fall short in terms of producing high-quality images and capturing texture details, particularly with pulmonary vessels. To address these issues, this paper proposes a new bone suppression framework, termed BS-Diff, that comprises a conditional diffusion model equipped with a U-Net architecture and a simple enhancement module to incorporate an autoencoder. Our proposed network cannot only generate soft tissue images with a high bone suppression rate but also possesses the capability to capture fine image details. Additionally, we compiled the largest dataset since 2010, including data from 120 patients with high-definition, high-resolution paired CXRs and soft tissue images collected by our affiliated hospital. Extensive experiments, comparative analyses, ablation studies, and clinical evaluations indicate that the proposed BS-Diff outperforms several bone-suppression models across multiple metrics. Our code can be accessed at \href{https://github.com/Benny0323/BS-Diff}{https://github.com/Benny0323/BS-Diff}.

\end{abstract}
\section{Introduction}
\label{sec:intro}
Morbidity and mortality rates associated with lung disease have remained high in recent years. Chest X-rays (CXRs) are a widely accepted low-dose technique for lung screening. Yet, even skilled clinicians can overlook lesions in lung areas that are not prominent, primarily due to the fact that 75\% of the lung area in CXRs overlaps with bone, thereby hindering disease detection and diagnosis. To enhance the accuracy of clinical diagnosis, a process known as bone suppression has been introduced. The most highly regarded method currently in use is the Dual-Energy Subtraction (DES) imaging technique. However, this technique necessitates costly equipment and exposes patients to elevated radiation doses. Consequently, researchers are endeavoring to identify less harmful and cost-effective bone suppression techniques.\par
In early methods, Suzuki et al. \cite{suzuki2006image} employed a massive artificial neural network to generate a bone image from CXRs. This bone image could then be subtracted to produce an image resembling a soft-tissue image. Subsequently, Juhasz et al. \cite{juhasz2010segmentation} used the active shape model to segment anatomical structures on CXRs and suppress bone shadows. They applied this model to the JSRT dataset \cite{shiraishi2000development}, which is currently the only publicly available dataset of its kind. However, these methods are easy to lack high-level semantic information pertaining to bony structures. Hence, several bone suppression methods have employed deep learning techniques to learn mappings from CXRs to soft tissues recently. Yang et al. \cite{yang2017cascade} developed a multi-scale convolutional neural network (CNN) model for bone suppression within the gradient domain of an image. This model could effectively learn sparse features and yield exceptional results. Gusarev et al. \cite{gusarev2017deep} viewed bones as noise levels by separately employing an autoencoder (AE) and a deep CNN with various loss functions to suppress bones. However, this approach resulted in blurry images due to its failure to capture high-frequency details. To reduce the blurriness, Zhou et al. \cite{zhou2019generation} proposed a multi-scale conditional generative adversarial network (GAN) to substantially preserve high-frequency details and increase the sharpness of the generated images. For improving the classification and detection accuracy of tuberculosis, Rajaraman et al. \cite{rajaraman2021chest} developed several models with the best-performing ResNet-BS to greatly suppress bones. Recently, Liu et al. \cite{liu2023bone} presented a bone suppression technique for lateral CXRs making use of a particular data rectification method and a distillation learning algorithm. Nevertheless, many of these works still suffer from issues such as harsh denoising or the generation of low-quality bone images, leading to a decrease in image sharpness and detailed texture \cite{fan2019brief}. Furthermore, the quality of the only currently accessible dataset is insufficient.\par
Recently, denoising diffusion probabilistic models (DDPMs) \cite{ho2020denoising}, a novel class of generative models, have surfaced to tackle issues of mode collapse and non-convergence prevalent in GAN. These models generate data by applying a series of transformations to random noise. In this paper, we present a new framework, named BS-Diff, integrating a conditional diffusion model (CDM) equipped with a U-Net architecture and a straightforward enhancement module employing an AE. Our proposed method has demonstrated its capability of generating high-quality images with a high degree of bone suppression, and its enhanced ability to capture intricate texture details, such as small pulmonary vessels. We have performed comprehensive experiments, comparisons, ablation studies, and clinical evaluations that collectively affirm the superiority of our BS-Diff over numerous existing bone-suppression models using multiple metrics. In addition, we have assembled the largest post-2010 dataset, comprised of high-quality, high-resolution data from 120 patients. The dataset, which includes paired CXRs and soft tissue images, was collected in collaboration with our partnering hospital. We expect to make this dataset publicly available in the near future. The key contributions of our study can be summarized as follows:
\begin{itemize} 
    \item To the best of our knowledge, this is the pioneering study that harnesses diffusion models for the generation of soft-tissue images from CXRs, thus addressing and overcoming the prevailing limitations of DES.
\end{itemize}
\begin{itemize}
    \item In our enhancement module, we introduce an innovative amalgamation of varied loss functions, devised to more effectively encapsulate spatial features and intricate texture details of images, while concurrently preserving their overall structures.
\end{itemize}
\begin{itemize}
    \item We have also assembled the most extensive dataset since 2010, comprising high-quality, high-resolution paired images from 120 patients, which were collected in collaboration with our partner hospital. We anticipate making this dataset publicly available in the near future.
\end{itemize}
\begin{itemize}
    \item Through comprehensive experiments, comparative analyses, ablation studies, and clinical evaluations, we substantiate the superior performance of our proposed BS-Diff model in comparison to several high-performing bone suppression models.
\end{itemize}

\section{Method}
\label{sec:format}

\subsection{Network architecture overview}
\begin{figure}[t]
\begin{minipage}[b]{1.0\linewidth}
  \centering
  \centerline{\includegraphics[width=8.5cm]{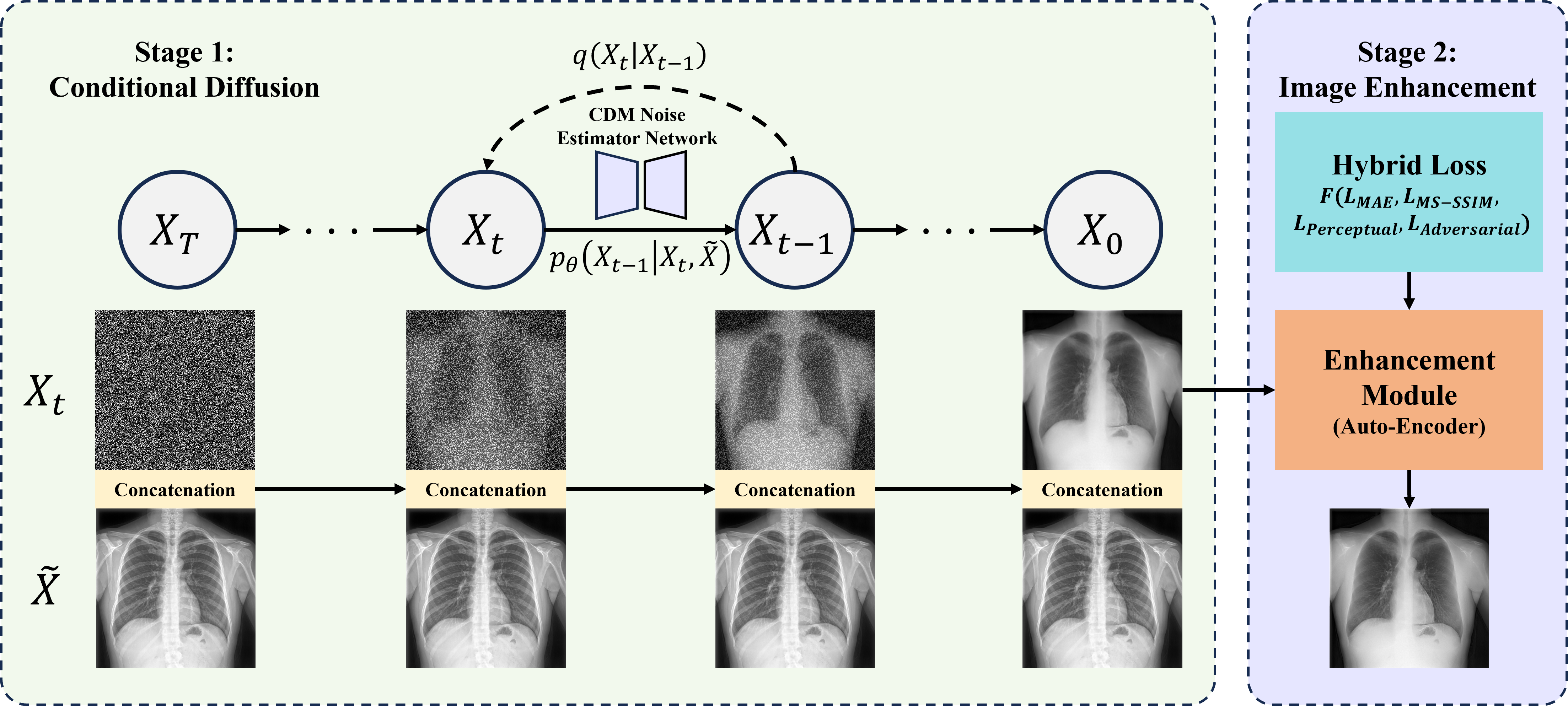}}
\end{minipage}
\vspace{-2em}
\caption{The overall architecture of our proposed BS-Diff.}
\label{fig:1}
\vspace{-0.5em}
\end{figure}

\vspace{-0.5em}
The overview of our proposed method is depicted in Fig.\ref{fig:1}. Our BS-Diff model typically operates in two stages: the first stage involves a CDM with a U-Net architecture while the second stage involves a straightforward enhancement module using an AE. In the first stage, the CDM generates estimated soft tissues taking concatenated Gaussian noise and CXRs as inputs. Subsequently, in the second stage, the AE uses the output of the CDM to generate sharper and higher-quality soft-tissue images. Crucially, our model should maintain the texture, color, and pulmonary vessels in the soft tissues, including those overlapping with bony structures.

\vspace{-0.5em}
\subsection{Conditional diffusion model}
\vspace{-0.5em}
Diffusion models comprise two components: a forward process and a reverse process. The forward noising process, denoted as $q$, is described as follows:
\vspace{-0.5em}
\begin{equation}
\vspace{-0.5em}
    q(x_t|x_{t-1}) = N(x_t; \sqrt{1-\beta_t}x_{t-1}, \beta_tI),
\end{equation}
where $\beta_t$ represents a predefined sequence of variances. 
The denoising process $p_\theta$ is derived through the optimization of the model parameters $\theta$, which can be defined as follows:
\vspace{-0.5em}
\begin{equation}
\vspace{-0.5em}
\begin{split}
    p_\theta(x_{t-1}|x_t) &= p(x_T)\prod\limits_{t=1}^T p_\theta(x_{t-1}| x_t) \\
&=N(x_{t-1}; \mu_\theta(x_t,t),\Sigma_\theta(x_t,t)).
\end{split}
\label{F7}
\end{equation}
Both $\mu_\theta$ and $\Sigma_\theta$ can be estimated using a U-Net-based network \cite{ronneberger2015u}, denoted as $\epsilon_\theta$, with parameters $\theta$. Then, the estimated $\mu_\theta$ and $\Sigma_\theta$ can be used to denoise images during sampling.
\par
The diffusion models mentioned above are unconditional, while a conditional one allows us to manipulate the generation results according to our specific intentions. The fundamental concept of our CDM is to learn a conditional reverse process $p_\theta(x_{1:T}|\widetilde{x})$, without altering the forward noising process, such that the sampled $x_0$ closely aligns with the data distribution conditioned on $\widetilde{x}$.\par
As is shown in Fig.\ref{fig:1}, during the training phase, we initially sample ($x, \widetilde{x}$) from a well-paired data distribution $q(x, \widetilde{x}$). This entails paired soft tissues $x$ with CXRs $\widetilde{x}$. Subsequently, we learn a noise predictor $\epsilon_\theta$, providing $\widetilde{x}$ as an input to the reverse process. This can be expressed as follows: \par
\begin{equation}
\vspace{-0.5em}
p_\theta(x_{0:T}|\widetilde{x}) = p(x_T)\prod\limits_{t=1}^T p_\theta(x_{t-1}| x_t, \widetilde{x}).
\end{equation}
Given that our model employs image-based conditioning, the previous formula of optimizing a noise predictor network as defined in formula (\ref{F7}) becomes $\mu_\theta(x_t, \widetilde{x}, t)$, $\Sigma_\theta(x_t, \widetilde{x}, t)$. In this case, the inputs $x$ and $\widetilde{x}$ are concatenated channel-wise, resulting in two-dimensional image channels. The U-Net architecture in our BS-Diff model largely mirrors that of the DDPM \cite{ho2020denoising}. Specifically, we defined five different levels and changed the number of residual blocks per level to 3. We also added the attention mechanisms in the last four levels, each with various channels per attention head.
\vspace{-1em}
\subsection{Enhancement module}
\vspace{-0.5em}
Though the diffusion model demonstrates strong generative capabilities in many computer vision tasks, it is apt to lose images' high-frequency information due to iterative noise added in the forward diffusion period. Therefore, to meet the high-definition demands of medical imaging, we introduce an enhancement module. This module is capable of transforming smoother images into high-quality ones, with particular emphasis on preserving fine information. To achieve this, we designed a straightforward AE model, which incorporates an encoder to transform the input data, and a decoder to recreate the input data from the encoded representation. The enhancement module is independently trained by utilizing its specific loss functions.
\vspace{-1em}
\subsection{Hybrid loss functions}
\vspace{-0.5em}
The conventional objective of the diffusion model, which aligns with our approach, is to predict noise applied to the image in the forward process using the mean squared error (MSE) loss. This prediction is accomplished through a noise estimator network. For the enhancement module, we propose a hybrid loss function aimed at aiding the network to generate clearer, sharper, and higher-quality images whilst maintaining superior texture details. Our proposed loss function incorporates multiple newly weighted losses. These losses include MAE loss, perceptual loss \cite{johnson2016perceptual}, via a pre-trained VGG-16 network, Multi-Scale Structural Similarity (MS-SSIM) loss \cite{wang2003multiscale}, and an adversarial loss on a patch discriminator based on the Pix2PixHD method \cite{wang2018pix2pixHD}. The final cost function can be expressed as follows:
\vspace{-0.5em}
\begin{equation}
\vspace{-0.5em}
\begin{split}
L & = \lambda_{mae} \cdot L_{MAE} + \lambda_{ms-ssim} \cdot L_{MS-SSIM} +  \lambda_{perceptual} \\
& \cdot L_{Perceptual} + \lambda_{adversarial} \cdot L_{Adversarial}.
\end{split}
\end{equation}
The specifications for the last three losses are detailed in the following sections.\par
\textbf{MS-SSIM Loss:} The MS-SSIM \cite{wang2003multiscale} is derived from modeling the luminance, contrast, and structure of an image across multiple scales. The formula is given as follows:
\begin{equation}
\vspace{-0.5em}
    MS-SSIM = [l_M(s_i, \widetilde{s}_i)]^{\alpha_M} \cdot \prod\limits_{j=1}^M [c_j(s_i, \widetilde{s}_i)]^{\beta_j} \cdot [s_j(s_i, \widetilde{s}_i)]^{\gamma_j} \\,
\end{equation}
\begin{equation}
\vspace{-0.5em}
    L_{MS-SSIM} = \frac{1}{N} \sum_{i=1}^N 1 - (MS-SSIM),
\end{equation}
where $l$, $c$, and $s$ represent luminance, contrast, and structure at varying scales denoted by '$j$', respectively. \par
\textbf{Perceptual Loss and Adversarial Loss:} The objective of the perceptual loss is to conserve the perceptual and semantic understanding of the image by employing deep features extracted from neural networks, specifically a pre-trained VGG-16 network. The adversarial loss, on the other hand, utilizes a patch discriminator based on a pre-trained pix2pixHD network \cite{wang2018pix2pixHD}. The patch discriminator's objective is to classify an image as either real or fake. For the sake of simplicity, both networks are denoted as $\phi$, owing to their identical expressions as follows:
\vspace{-0.75em}
\begin{equation}
\vspace{-0.25em}
    L_{Perceptual/Adversarial}= \frac{1}{N} \sum_{i=1}^N (\phi(s_i) - \phi(\widetilde{s}_i)).
\end{equation}
\par
\section{EXPERIMENTAL RESULTS}
\label{sec:typestyle}
\subsection{Data preparation and preprocess}
\vspace{-0.5em}
We gathered a collection of 163 paired posterior-anterior DES CXRs, shot by a digital radiography (DR) MCAhine equipped with a two-exposure DES unit (Discovery XR656, GE Healthcare), from our partnering hospital. The images were initially stored in DICOM format with a 14-bit depth but were later converted to PNG files for convenience. All CXRs have 2021 × 2021 pixel dimensions and a pixel size range of 0 to 0.1943 mm. We excluded 43 paired radiographs including operational errors, significant motion artifacts and visible pleural effusions, pneumothorax. The final dataset for the experiments consisted of 120 paired images. To enhance our model's stability and convergence rate, we applied various data augmentation methods, including horizontal flip, image rotation and various contrast adjustment techniques, resulting in a total of 840 paired images post-augmentation. We split the whole dataset into training, validation and testing, and the ratio is 7:2:1. All images were resized to 256 × 256 pixels to conserve memory. We also employed image registration to optimize the fusion of information between paired images and contrast-limited adaptive histogram equalization for local contrast enhancement. All image pixel values were later normalized to the range [-1, 1].
\vspace{-1em}
\subsection{Implementation details}
\vspace{-0.5em}
All experiments were executed using the PyTorch 2.0.1 framework on a single Nvidia A100 GPU. Both stages of our model were trained from scratch for 200 epochs with a batch size of 2. The parameters were optimized using the AdamW optimizer algorithm. An exponential moving average with a rate of 0.995 is applied to the parameters. We also incorporated a dynamic learning rate schedule for both stages, starting with an initial learning rate of 0.0001. For our CDM, the training and sampling steps $T$ were set to 1000. The parameter $\beta$ ranged from 0.0015 to 0.0205, mapping to $T$ steps using a cosine schedule. For our enhancement module, the weights were as follows: $\lambda_{mae}=1.0, \lambda_{ms-ssim}=1.0, \lambda_{perceptual}=0.001, \lambda_{adversarial}=0.01$.
\vspace{-1em}
\subsection{Comparisons}
\begin{figure}[t]
\begin{minipage}[b]{1.0\linewidth}
  \centering
  \centerline{\includegraphics[width=8.5cm]{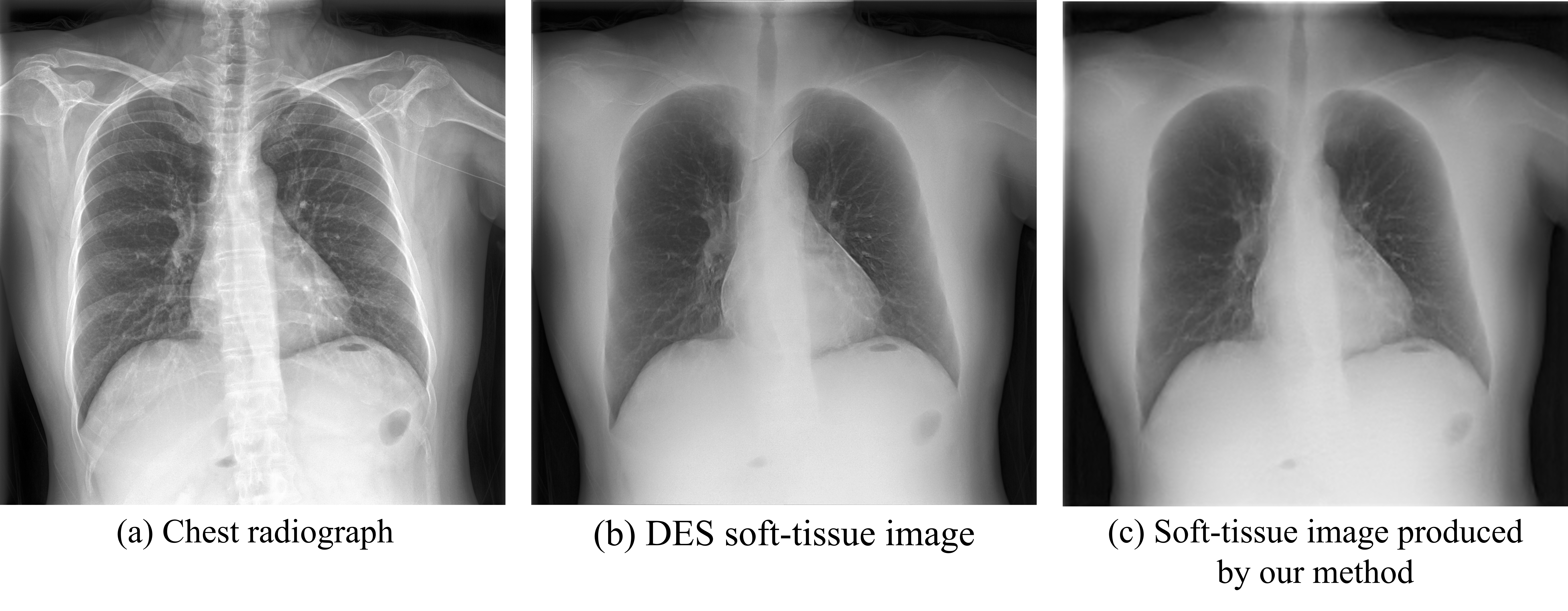}}
\end{minipage}
\vspace{-2.5em}
\caption{Visualization of CXR, DES soft-tissue image and soft-tissue image produced by our method.}
\label{fig:2}
\vspace{-1.5em}
\end{figure}

\vspace{-0.5em}
Based on the principle of fair comparison, we compared the proposed model with three methods from previous works: the multi-scale conditional adversarial network (MCA-Net) \cite{zhou2019generation}, an autoencoder-like convolutional model \cite{gusarev2017deep}, and the ResNet-BS model \cite{rajaraman2021chest}. To evaluate the generated soft-tissues, we employed bone suppression ratio (BSR) \cite{yang2017cascade} ($\uparrow$), SSIM ($\uparrow$), MSE ($\downarrow$), and PSNR ($\uparrow$) metrics. Table \ref{T1} illustrates the performance of different methods. Overall, our method outperforms the others across all metrics. The ResNet-BS records the poorest performance based on all evaluation metrics due to luminance differences and significant loss of texture, while the autoencoder-like convolutional model and MCA-Net show comparable results. Broadly, our method surpasses the competing models and improves BSR, MSE, SSIM, and PSNR by a minimum of 1.7\%, 0.003, 0.015, and 0.164, respectively. The results of our proposed BS-Diff with two stages are demonstrated in Fig. \ref{fig:2}, which captures intricate image details, preserves overall structures, and enhances performance. 

\begin{table}[t]
\caption{Comparison performance with previous works.\\}
\centering
\begin{tabular}{|c|c|c|c|c|}
\hline
Methods & BSR  & MSE    & SSIM  & PSNR   \\ \hline
MCA-Net \cite{zhou2019generation} & 0.971    & 0.0005  & 0.922 & 38.085 \\ \hline
Gusarev et al. \cite{gusarev2017deep}      & 0.960    & 0.0007 & 0.885 & 34.296 \\ \hline
ResNet-BS \cite{rajaraman2021chest}  & 0.725    & 0.0041 & 0.823 & 29.063 \\ \hline
\textbf{BS-Diff (Ours)} & \textbf{0.988}    & \textbf{0.0002} & \textbf{0.937} & \textbf{38.249} \\ \hline
\end{tabular}
\label{T1}
\vspace{-1.5em}
\end{table}

\vspace{-1em}
\subsection{Ablation studies}
\vspace{-0.5em}
To evaluate the significance of the enhancement module proposed in BS-Diff, our model was trained both with and without this module. We found that exclusive reliance on the CDM produced smoother images, omitting some textural details; this was not the case with images produced with the enhancement module. It is evident that the employment of the enhancement module results in superior, especially sharper images, and improved both the PSNR and BSR scores by 8.786 and 0.153 respectively, as demonstrated in Table \ref{T2}.
\vspace{-1em}

\begin{table}[t]
\vspace{-0.5em}
\caption{Ablation performance with and without enhancement module.\\}
\vspace{-1em}
\centering
\begin{tabular}{|c|c|c|c|c|}
\hline
Methods & BSR  & MSE    & SSIM  & PSNR   \\ \hline
BS-Diff w/o AE & 0.796   & 0.0029 & 0.893 & 29.528 \\ \hline
\textbf{BS-Diff w/AE} & \textbf{0.988}    & \textbf{0.0002} & \textbf{0.937} & \textbf{38.249} \\ \hline
\end{tabular}
\label{T2}
\vspace{-1.5em}
\end{table}

\begin{table}[t]
\caption{Clinical evaluation for our proposed BS-Diff.\\}
\centering
\begin{tabular}{|cl|c|}
\hline
\multicolumn{2}{|c|}{\makecell[c]{Clinical Evaluation Criteria}} & \multicolumn{1}{l|}{Score}\\ \hline
\multicolumn{1}{|c|}{\multirow{3}{*}{\makecell[c]{Pulmonary\\ vessels\\visibility}}}& \makecell[c]{Clearly displayed (3)}
& \multirow{3}{*}{2.67}
\\ \cline{2-2}
\multicolumn{1}{|c|}{}&\makecell[c]{ Displayed (2)}
&\\ \cline{2-2}
\multicolumn{1}{|c|}{}& \makecell[c]{Not displayed (1)}
&\\ \hline
\multicolumn{1}{|c|}{\multirow{3}{*}{\makecell[c]{Central\\ airway\\ visibility}}}& \makecell[c]{Lobar and intermediate bronchi (3)}
& \multirow{3}{*}{2.33}\\ \cline{2-2}
\multicolumn{1}{|c|}{}& \makecell[c]{Main bronchus and rump (2)}&\\ \cline{2-2}
\multicolumn{1}{|c|}{}& \makecell[c]{Trachea (1)}
&\\ \hline
\multicolumn{1}{|c|}{\multirow{3}{*}{\makecell[c]{Degree of\\ bone sup-\\pression}}}& \makecell[c]{Nearly perfect suppression (3)}
& \multirow{3}{*}{2.33}
\\ \cline{2-2}
\multicolumn{1}{|c|}{}& \makecell[c]{Unsuppressed bones less than 5 (2)}
&\\ \cline{2-2}
\multicolumn{1}{|c|}{}& \makecell[c]{5 or more bones unsuppressed (1)}
&\\ \hline
\end{tabular}
\label{T3}
\vspace{-1.5em}
\end{table}

\subsection{Clinical evaluation}
\vspace{-0.5em}
The bone suppression images produced by our model were independently evaluated by three recruited physicians of varying expertise in our collaborative hospital, in accordance with the bone suppression application evaluation criteria. The average scores (maximum: 3) are shown in Table \ref{T3}. Our results demonstrated that our soft-tissues can clearly preserve visibility of pulmonary vessels and central airways and greatly suppress bones, which can significantly improve the clinician's performance in finding lung lesions.
\vspace{-0.5em}


\section{Conclusions}
\vspace{-0.5em}
\label{sec:majhead}
To mitigate the high costs and dose concerns associated with DES devices, this paper introduces a novel bone suppression framework, BS-Diff, comprising a CDM with a U-Net architecture and a straightforward enhancement module incorporating an AE. Our method is capable of producing high-quality images with a high bone suppression rate and amplifying the ability to discern fine texture information. Comprehensive experiments and clinical evaluations demonstrate that our proposed BS-Diff outperforms multiple existing bone suppression models across several metrics. Also, we have compiled the largest dataset after 2010, which includes data from 120 patients with paired CXRs and soft tissue images gathered from our partner hospital. But there is still room for improvement. We don't perform downstream tasks like detection or classification in this paper. Improving our model's architecture or introducing better condition-guided methods to precisely control generation could be explored further.
\section{Compliance with Ethical Standards}
\vspace{-0.5em}
This research study was conducted retrospectively and in accordance with the principles of the Declaration of Helsinki. We applied for an informed consent waiver. Approval was granted by the Ethics Committee of Cancer Hospital \& Shenzhen Hospital, Chinese Academy of Medical Sciences and Peking Union Medical College (JS2023-19-2).
\section{Acknowledgements}
This work was supported by the Zhejiang Provincial Natural Science Foundation of China (No.LY21F020017, 2022C03043), GuangDong Basic and Applied Basic Research Foundation (No.2022A1515110570), Shenzhen Science and Technology Program (No.KCXFZ20201221173008022), Innovation Teams of Youth Innovation in Science and Technology of High Education Institutions of Shandong Province (No. 2021KJ088). All authors declare that they have no conflicts of interest.

\bibliographystyle{IEEEbib}
\bibliography{refs_simple}
\end{document}